\documentclass[aps,prl,twocolumn,superscriptaddress]{revtex4}

\usepackage{graphicx}
\usepackage{bm}

\begin{document}






\date{\today}

\title{Electron Heating in Quasi-Perpendicular Shocks}
\author{F.~S.~Mozer}
\email{forrest.mozer@gmail.com}
\affiliation{Space Sciences Laboratory, University of California, Berkeley, CA. 94720 USA}
\author{D.~Sundkvist}
\affiliation{Space Sciences Laboratory, University of California, Berkeley, CA. 94720 USA}
\date{\today }

\begin{abstract}
Seventy crossings of the Earth’s bow shock by the THEMIS satellites have been used to study thermal electron heating in collisionless, quasi-perpendicular shocks. It was found that the temperature increase of thermal electrons differed from the magnetic field increase by factors as great as three, that the parallel electron temperature increase was not produced by parallel electric fields, and that the parallel and perpendicular electron temperature increases were the same on the average. It was also found that the perpendicular and parallel electron heating occurred simultaneously so that the isotropization time is the same as the heating time. These results cannot be explained by energy transfer from waves to electrons or by the motion of magnetized electrons through the shock. Electric field fluctuations on the scale of the electron gyro-diameter were found to be  of finite amplitude in the shock ramp, which requires that the electron trajectories be more random and chaotic than orderly and adiabatic. The data may be explained by the large amplitude electric field fluctuations that demagnetize electrons as they move through the cross-shock electric field. 
\end{abstract}


\maketitle

\section{Introduction}
During the half-century since the prediction \cite{axford1962a,kellogg1962a} and measurement \cite{ness1964a} of collisionless shocks, considerable study on mechanisms for converting supersonic plasma flow into thermal energy 
of plasma, fields, waves and accelerated particles has been expended. 
The removal of flow energy through dissipation, with the addition of dispersion, is necessary to limit the nonlinear steepening of the shock \cite{kennel1985a,treumann2009a}.

The heating mechanisms for ions and electrons are believed to be very different. While ion heating is fairly well understood, electron heating remains controversial. Proposed mechanisms for electron heating include 
(see e.g.  \cite{goodrich1984a,papadopoulos1985a,schwartz1988a,balikhin1993a,balikhin1994a,gedalin1995a,scudder1986a,scudder1986b,scudder1986c,schwartz2011a}):
\begin{itemize}
\item 
Adiabatic heating of magnetized electrons. The source of heating is the compression of the magnetic field across the shock ramp. 
This reversible mechanism predicts heating perpendicular to the magnetic field ($T_{e\perp}$). 
\item
Kinematic heating of magnetized electrons, where the cross-shock potential in the ramp gives a correction to the adiabatic moment. This scenario is also dependent on magnetic compression, and leads to super-adiabatic perpendicular heating ($T_{e\perp}$).
\item 
Parallel heating ($T_{e\|}$) of magnetized electrons by the parallel electric field in the shock ramp.
\item 
Energy and momentum transfer from waves to electrons through processes such as anomalous resistivity. 
Here the source is unstable currents or particle distributions in the shock layer giving rise to e.g. lower-hybrid, ion-acoustic or whistler micro-instabilities and turbulence. Wave-particle interaction causes momentum exchange, where waves gives energy to the particles (anomalous resistive heating).
\end{itemize}
Of these mechanisms the first three are due to macroscopic fields in the shock, while the latter is microscopic in nature.

Through examination of 70 terrestrial bow shock crossings made by the THEMIS satellites, it is shown 
in this paper 
that none of these mechanisms adequately explain the electron heating observations. Instead, it will be shown that electron heating is achieved by turbulent transport of unmagnetized electrons through the cross-shock electric field.

The paper is organized as follows. First the instrumentation and data sets are presented.
Then, the above-listed heating mechanisms are investigated, one by one, and it is shown that none of them are consistent with the experimental data.  Finally, the model of electric field turbulence that demagnetizes electrons as they move through the cross-shock electric field is found to be adequate to explain the data.

\section{Instrumentation and Data Sets}

We use data from the fluxgate magnetometer \cite{auster2008a}, the electron electrostatic analyzer \cite{mcfadden2008a}, and the electric field instrument \cite{bonnell2008a} on the five THEMIS spacecraft to study bow shock crossings at times when the data transmission rate was 128 or 8192 samples per second and, usually, when three of the spacecraft encountered the Earth's bow shock within a short time interval. We selected quasi-perpendicular shocks, all of which were super-critical with respect to the Alfvenic Mach number, $M_A$.

\begin{figure}
\includegraphics[width=\columnwidth]{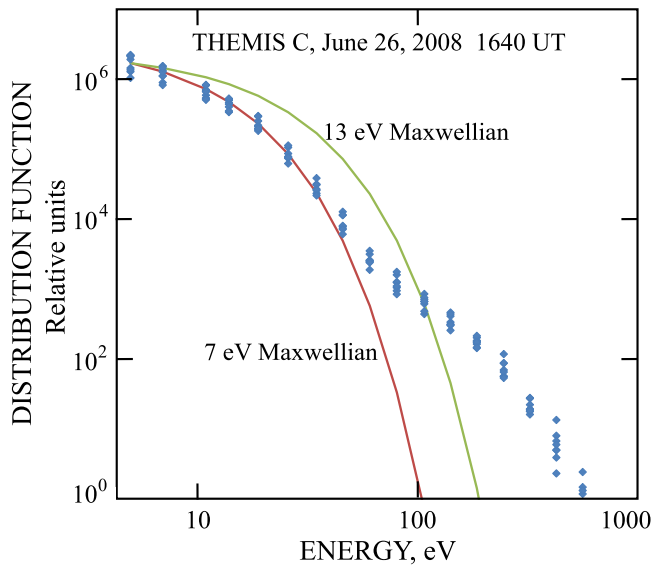}%
\caption{\label{fig:maxwelltemperatures}Distribution function of $90^\circ$ pitch angle electrons measured during a one second interval in the solar wind, and Maxwell-Boltzmann curves for temperatures of 7~eV and 13~eV (13~eV is the moments estimate of the temperature for this data).}
\end{figure}

Electron temperatures are conventionally obtained from moments of the electron distribution function. In this study we, instead, use the electron temperature estimated from fitting Maxwell-Boltzmann curves to the electron distributions, as illustrated in Figure~\ref{fig:maxwelltemperatures}. 
In this Figure, the blue diamonds are perpendicular ($75^\circ$ to $105^\circ$) pitch angle electrons measured during a one-second-interval upstream of the Earth's bow shock. This distribution consists of a low energy thermal core and a supra-thermal power law of halo electrons. The Maxwell-Boltzmann fit to the thermal core is the red curve characterized by a 7~eV temperature. The moments-temperature of 13~eV is illustrated by the green Maxwellian. The moments-temperature is greater than the Maxwellian-temperature because the moments fit includes contributions from the supra-thermal electrons. In the current data set, the moments-temperature exceeds the Maxwellian temperature by factors between about 1.2 and 2.0. An additional reason for using Maxwellian-temperatures in this study is that they may be computed at a cadence much less than one second, while the moments-temperature is available only once every 
spin period (four seconds for the THEMIS spacecraft).

\section{Heating by Energy Transfer From Waves to the Electrons does not Explain the Data}
The idea that anomalous resistivity can be a channel for heating of plasma in shocks has been studied since the early days of collisionless shock research. In this scenario unstable currents in the shock transition layer, as well as unstable particle distributions and gradients in density, temperature and magnetic field, can drive instabilities which excite a spectrum of waves. These waves would then, through wave-particle interaction, lead to momentum and energy exchange between waves and particles. While this process is microscopic (kinetic) the macroscopic impact is equivalent to a form of resistivity with an associated collision frequency. Since there are no binary collisions taking place, the resulting resistivity is called "anomalous". 
A formal derivation using the Vlasov equation led to an expression for the effective collision frequency \cite{silin2005a}.
A similar derivation from a generalized Ohm's law \cite{che2011a,mozer2011a}, leads to the expression for the part of the cross-shock electric field due to electrostatic anomalous drag or resistivity
\begin{equation}\label{eq:drag}
D_x = -\frac{<\delta n \delta E_x>}{<n>}
\end{equation}
where $n$ is the plasma density and $E_x$ is the electric field. $D_x$ provides the dissipation required for converting electromagnetic energy to plasma energy and it has been used to show that wave heating in magnetic field reconnection is unimportant \cite{mozer2011a}. 

\begin{figure}
\includegraphics[width=\columnwidth]{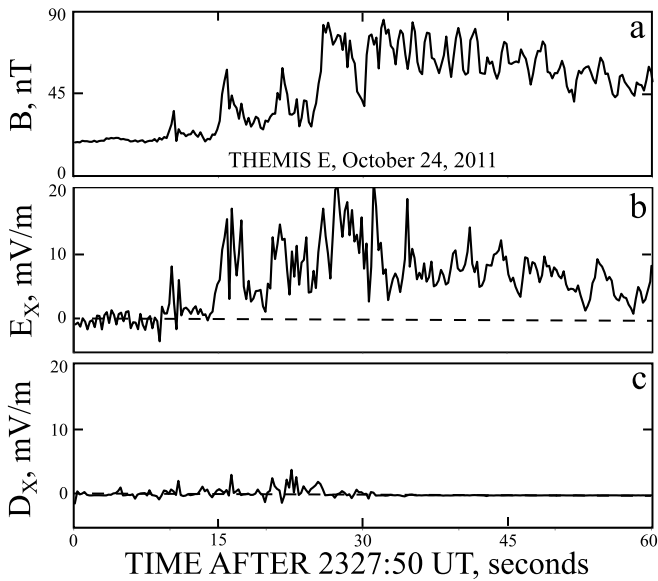}%
\caption{\label{fig:drag}Measurements across a bow shock on October 24, 2011.  Panel 2a gives the magnetic field magnitude, panel 2b presents the cross-shock electric field, and panel 2c gives the anomalous drag, D.  Because D is much less than the cross-shock electric field, conversion of wave energy to particle heating is relatively unimportant.}
\end{figure}

In Equation~\ref{eq:drag}, the numerator includes the fluctuations of the electric field and density and their correlation. It is computed for a bow shock crossing on October 24, 2011 in Figure~\ref{fig:drag} (this crossing is discussed in greater detail in figures~\ref{fig:X},\ref{fig:XX} and \ref{fig:NIF}). Panel \ref{fig:drag}a of this figure gives the 0.25~s average of the total magnetic field during the crossing, panel \ref{fig:drag}b gives the 0.25~s average of the cross-shock electric field, and panel \ref{fig:drag}c gives the 0.25~s average of $D_x$. Because $D_x \ll E_x$, wave heating of electrons across the bow shock is unimportant. 
Although the fluctucations in the electric field are large compared to the mean field (see Figure~\ref{fig:NIF}d), and the density fluctuations are large compared to the mean density (see Figure~\ref{fig:X}b), $D_x$ is small because the electric field and density fluctuations are uncorrelated.
\\

\section{Perpendicular Heating of Magnetized Electrons does not Explain the Data}
Early work by Goodrich and Scudder \cite{goodrich1984a,scudder1986a,scudder1986b,scudder1986c,scudder1995a} considered magnetized electrons.
To a first approximation, magnetized electrons conserve the fluid first adiabatic invariant, $T_{e\perp}/B$, across the shock ramp 
where $T_{e\perp}$ is the perpendicular Maxwellian-temperature and B is the magnetic field magnitude.

\begin{figure}
\includegraphics[width=\columnwidth]{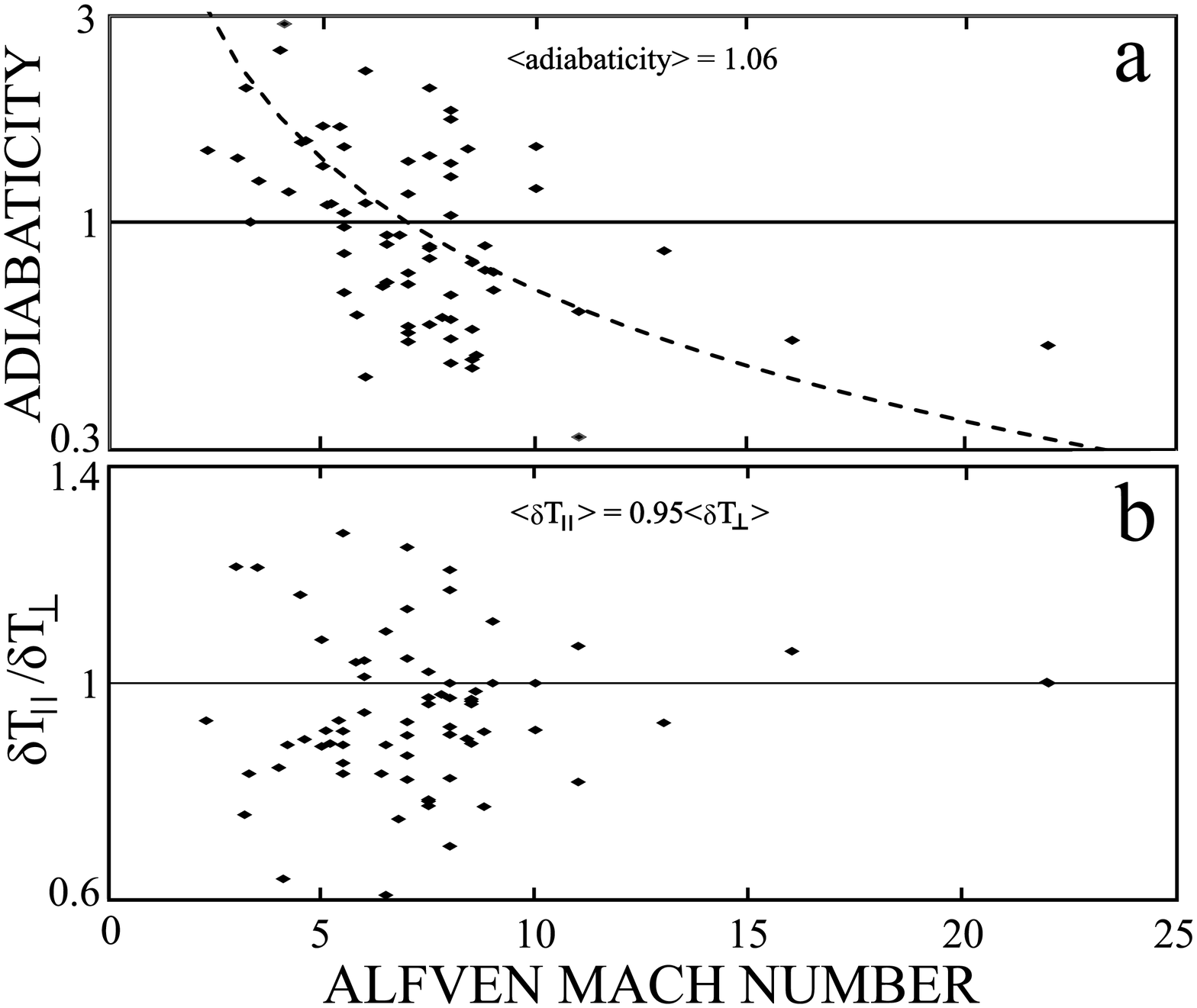}%
\caption{\label{fig:adiabaticity}Scatter plots of the adiabaticity (panel 3a) and ratio of parallel to perpendicular electron heating (panel 3b) for 70 bow shock crossings.   Panel 3a shows that the perpendicular electron heating was as much as three times larger or smaller than that expected for adiabatic compression.  Panel 3b shows that the average parallel and perpendicular electron heating were the same.}
\end{figure}

The adiabaticity is defined as
\begin{equation}
\mathrm{adiabaticity} = (T_{e\perp,sh}/B_{sh})/(T_{e\perp,sw}/B_{sw}) 
\end{equation}
where $sw$ and $sh$ refer to the upstream (solar wind) and downstream (sheath) parameters. For magnetized electrons the adiabaticity should be equal to one. For 70 quasi-perpendicular ($\theta_{Bn}$ between $46^\circ$ and $89.5^\circ$) shocks, the adiabaticity is plotted in Figure~\ref{fig:adiabaticity}a versus the Alfv\'en Mach number, $M_A$. 
At individual crossings, thermal electrons may be heated more or less than adiabatically by factors as great as three (although the average adiabaticity is $1.06 \pm 0.52$). 
Thus, the model of 
conservation of the first adiabatic invariant used to explain heating in the compressed magnetic field is inconsistent with this experimental data.
The adiabaticity obtained using moments-temperatures has the same characteristic as does the data of Figure~(\ref{fig:adiabaticity})a.
\section{Electron HEATING in parallel electric fields does not explain the data}
\begin{figure}
\includegraphics[width=\columnwidth]{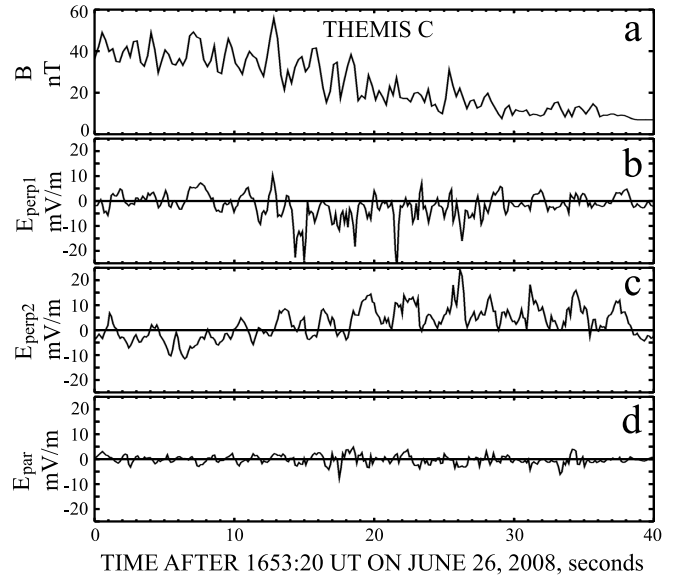}%
\caption{\label{fig:parallel_electric_field}The total magnetic field (panel 4a) across a bow shock and the three components of the electric field in the deHoffman-Teller frame, which is the frame in which the perpendicular electric field is zero in the solar wind (at the right of the plot).  The fields are in a magnetic-field-aligned coordinate system with the electric field component parallel to B given in panel 4d.  In this example, and for all of the 70 events, the parallel electric field in the shock ramp was small compared to the perpendicular field in this or any fixed coordinate system. }
\end{figure}

Parallel heating of magnetized electrons cannot be achieved by adiabatic compression. Instead we consider parallel heating by the parallel electric field. 
In the deHoffman-Teller (HT) frame \cite{deHoffmann1950a}, the perpendicular electric field in the solar wind is zero, 
so the plasma moves parallel to the magnetic field. In an idealization, the perpendicular electric field is also zero through the shock, and downstream. 
In this case, the parallel heating can only be achieved by the parallel electric field in the ramp. To test this idealized model, a typical example of the electric field at a shock crossing is given in Figure~(\ref{fig:parallel_electric_field}). 
Panel \ref{fig:parallel_electric_field}a gives the total magnetic field and the remaining panels give the HT electric field in a magnetic-field-aligned coordinate system in which panel \ref{fig:parallel_electric_field}d is the component parallel to the instantaneous magnetic field. The electric field components in the HT-frame are all small upstream and downstream. The parallel component of panel \ref{fig:parallel_electric_field}d is zero to within the experimental uncertainties and, contrary to the idealized model, the perpendicular components in panels \ref{fig:parallel_electric_field}b and \ref{fig:parallel_electric_field}c are large in the ramp. 
It is noted that, for the event of Figure~\ref{fig:parallel_electric_field}, the magnetic field was nearly in the spacecraft spin plane so the parallel electric field was largely measured by the long spin-plane electric field sensors and the large uncertainties associated with the short spin-axis electric field sensor had only a small effect in the parallel field measurement.

This data is representative of our observations that, except for occasional short duration bipolar parallel fields due to Debye scale electron holes, the parallel electric field in the shock ramp is much smaller than the perpendicular field. 
Thus, the parallel electron heating does not arise from the parallel electric field in the ramp, but rather, from the effects of the perpendicular electric field.

This fact is confirmed in Figure~\ref{fig:adiabaticity}b where it is shown that the parallel and perpendicular electron heating in each of the 70 events differed by at most 40\% and was, on the average, equal. 

\section{Unmagnetized Electrons}
Electrons would be unmagnetized if fluctuations of the electric field over the electron gyro-orbit were significant. To test if this is the case, fluctuations, $\delta \phi$, in the cross-shock potential 
(where $\phi = \int E_x v_{sh} dt$ and $v_{sh}=10~km/s$) as well 
as density fluctuations, $\delta n$, are used to compute
\begin{equation}
\delta n /n = [n(t+\tau) - n(t)]/<n>
\end{equation}
\begin{equation}
\delta \phi /T_e = [\phi(t+\tau) - \phi(t)]/<T_e>
\end{equation}
where $\tau$ is the time required for the shock to move a distance across the spacecraft equal to the electron gyro-diameter and the averages in the denominators are computed over the time interval from $t$ to $(t+\tau)$. The electrons are unmagnetized if 
$\delta \phi/T_e$ is of
the order of one. 
\begin{figure}
\includegraphics[width=\columnwidth]{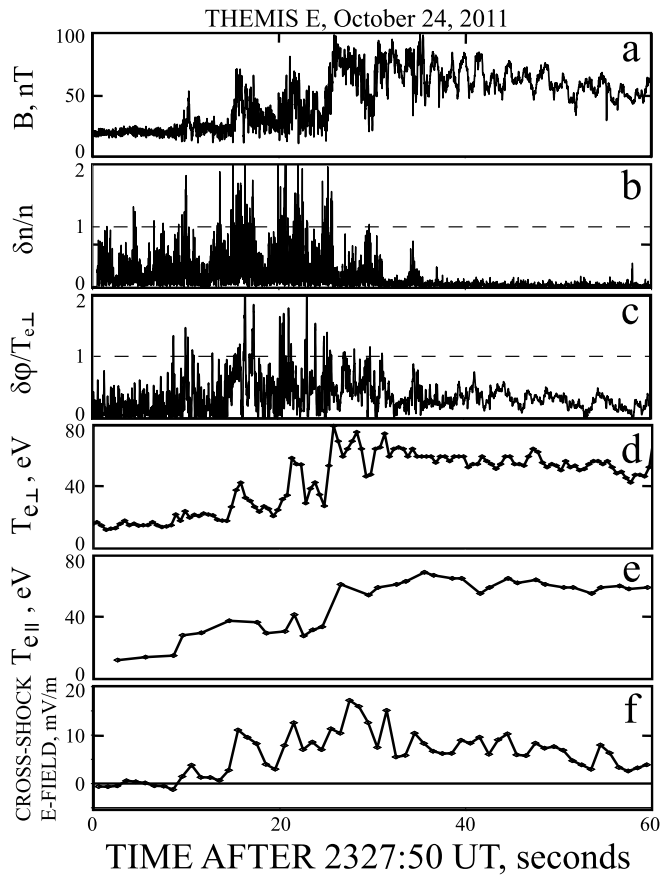}%
\caption{\label{fig:X}
The bow shock crossing that is also illustrated in Figures~\ref{fig:drag}, \ref{fig:XX}, and \ref{fig:NIF}. This data is in the normal incidence frame (NIF) in which fluctuations over the electron gyro-diameter of the plasma density (panel \ref{fig:X}b) and the cross-shock electric potential (panel \ref{fig:X}c) are important in the shock ramp.  The perpendicular electron heating (panel \ref{fig:X}d) correlates with the electric field turbulence of panel \ref{fig:X}c and the one-second-averaged electric field of panel \ref{fig:X}f.  The parallel heating (panel \ref{fig:X}e), has the same amplitude and time dependence as does the perpendicular heating (within the 1-2 second time resolution of the parallel measurement).  This indicates that the parallel and perpendicular heating result from the same physical process.  Their correlation with the potential fluctuations of panel \ref{fig:X}c and the electric field of panel \ref{fig:X}f shows that this process involves unmagnetized electrons moving in the cross-shock electric field.
}
\end{figure}
Figure~\ref{fig:X} presents data for a shock crossing by THEMIS E on October 24, 2011 (the same event as in Figure~\ref{fig:drag}). 
The field data was obtained with a sampling rate of 128~S/s, and the perpendicular temperature was measured every 0.25~s.
The parameters for this shock were $M_A=5.5$, $\theta_{Bn}=85^\circ$, the normal was $n_{\mathrm{dsl}}=(0.9991, -0.0235, 0.0356)$ in despun spacecraft coordinates (essentially Geocentric Solar Ecliptic coordinates), and the local time of the crossing was 14:30~UT. Panel \ref{fig:X}a gives the total magnetic field, panel \ref{fig:X}b gives $\delta n/n$ and panel \ref{fig:X}c gives $\delta \phi/T_{e\perp}$. The quantity $\tau$ in these computations is 70~msec, which is the time required for the shock to move the gyrodiameter of a 40~eV electron in a magnetic field of 50~nT. Through the shock ramp, the gyrodiameter changed by a factor of about two, so the computations of $\delta n/n$ and $\delta \phi/T_e$ were also done for $\tau$ values ranging over a factor of two
with similar results to those presented in Figure~\ref{fig:X}. These results are that $\delta \phi/T_{e\perp}$ is of the order of one in several locations in the shock ramp, so the electron trajectories must be more random and chaotic than orderly and adiabatic, and the thermal heating of electrons must be of a stochastic nature.

\begin{figure}
\includegraphics[width=\columnwidth]{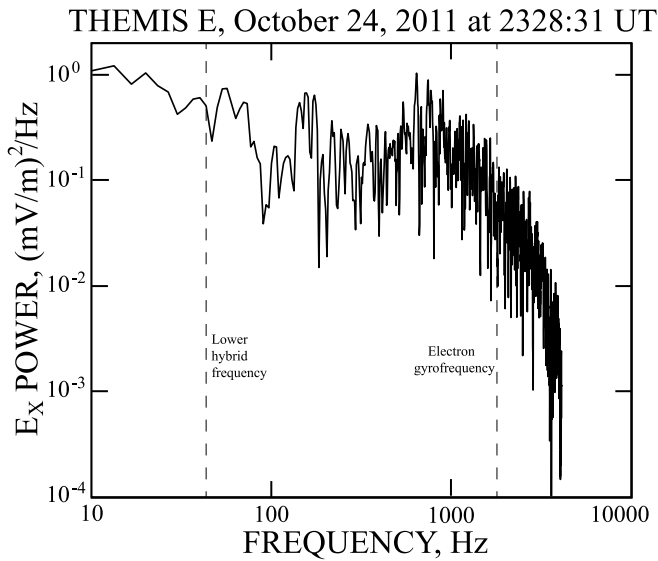}%
\caption{\label{fig:XX}Spectrum of the electric field fluctuations in the bow shock crossing of Figure~\ref{fig:X}.  Although there is power at and below the lower hybrid frequency, the dominant wave power is in the whistler frequency range.
}
\end{figure}

Data were collected at 128~samples/second through the entire crossing of Figure~\ref{fig:X}, corresponding to observations of fluctuations below 50~Hz. For a period of 0.3~s in the ramp, data were also collected at 8192~samples/second. The power spectrum of the cross-shock electric field for this time interval is given in Figure~\ref{fig:XX}. 
Fluctuations near the lower hybrid frequency and in the whistler mode range were observed, with the average waveform amplitude in the whistler mode range being a factor of 2.6 greater than the average amplitude covered by the 128 Hz sampling rate. Thus, the 128 Hz fluctuations, were multiplied by 3.6 to produce the $\delta \phi/T_{e\perp}$ plot of Figure~\ref{fig:X}c.
These fluctuations in $\delta \phi/T_{e\perp}$ are not bipolar electrostatic structures \cite{bale2002b} at Debye scales in the ramp. 
Such stuctures are observed occasionally in this and other shock crossings, but their integrated potentials are small \cite{bale2007a}.  

The data of Figure~\ref{fig:X} are in the Normal Incidence Frame (NIF). The NIF is defined as a shock stationary frame where the incident plasma flow is aligned along the shock normal.
\begin{figure}
\includegraphics[width=\columnwidth]{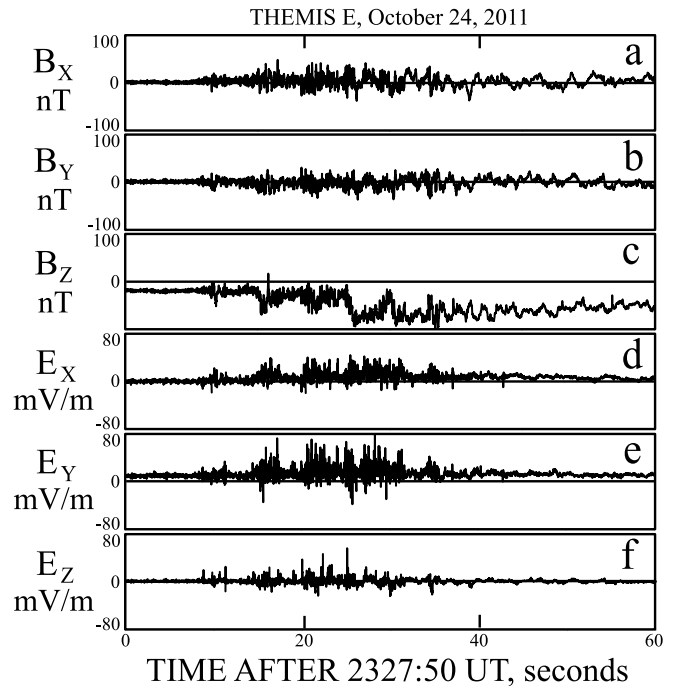}%
\caption{\label{fig:NIF}The three components of the electric and magnetic field during the bow shock crossing illustrated in the previous figures.  The data are in the normal incidence frame (NIF) such that the solar wind plasma flow is in the direction of the shock normal, X. The magnetic field component of panel \ref{fig:NIF}c and the normal, determined from a minimum variance analysis, are in the co-planarity plane, and the cross-shock electric field (panel d) is small upstream but positive in the ramp and downstream.}
\end{figure}
The fields in this frame are given in Figure~\ref{fig:NIF}, in which the top three panels are the components of $\mathbf{B}$ and the bottom three panels are the components of $\mathbf{E}$. The x-direction is normal to the shock. The co-planarity plane includes $B_z$ and $E_x$. In the solar wind, to the left of the plot, $E_x\sim 0$ (panel \ref{fig:NIF}d), while in the ramp and later it is clearly positive. Thus, there is a cross-shock electric field and a net potential. The combination of finite amplitude electric field fluctuations and associated potential in the presence of the DC cross-shock electric field provides the mechanism for heating the electrons as the following discussion illustrates.

Figure~\ref{fig:X}d gives the perpendicular Maxwellian-temperature, Figure~\ref{fig:X}e gives the parallel Maxwellian-temperature, and  Figure~\ref{fig:X}f gives the 1~s averaged cross-shock electric field.
The correlation between the perpendicular heating (panel \ref{fig:X}d), the electric field fluctuations (panel \ref{fig:X}c) and the cross-shock DC electric field (panel \ref{fig:X}f) is remarkable. This shows that strong electric field fluctuations of finite order, in association with the non-zero cross-shock electric field causes the electron heating in the shock ramp. 
In the latter portion of the plot, there continues to be a positive electric field
(panel \ref{fig:X}f) although the electron temperatures (panels \ref{fig:X}d and \ref{fig:X}e) do not increase. 
Thus, in the absence of turbulence, the perpendicular electric field caused nothing more than an $\mathbf{E}\times \mathbf{B}/B^2$ drift.  
The parallel electric field, being  small or zero, also did not heat electrons at this time.

The parallel electron temperature, obtained for electrons with pitch angles between $0^\circ$ and $30^\circ$, is shown in Figure~\ref{fig:X}e.  Within the 1-2 second time resolution of this measurement, the parallel and perpendicular temperatures were equal and their rise times were the same.  This indicates that the parallel and perpendicular heating result from the same physical process.  Their correlation with the potential fluctuations of panel \ref{fig:X}c shows that this process involves unmagnetized electrons in the cross-shock electric field of Figure~\ref{fig:NIF}d.

\begin{figure}
\includegraphics[width=\columnwidth]{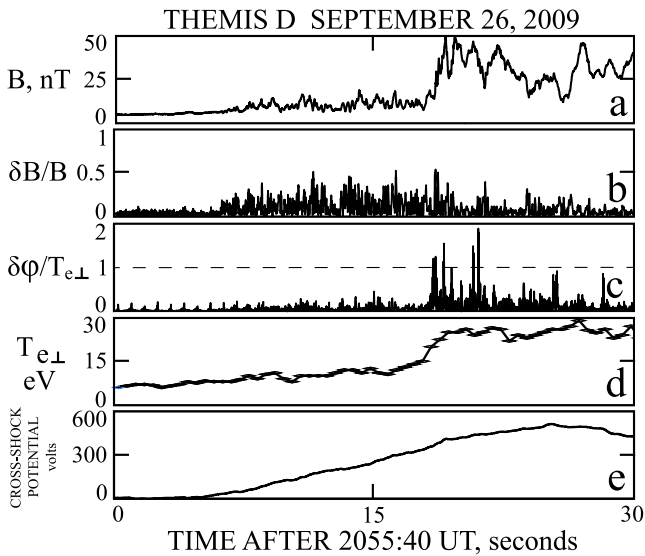}%
\caption{\label{fig:extra}Example of a bow shock crossing during which perpendicular electrons were heated in the shock ramp (panel d) coincident with large electric potential fluctuations (panel c) and the magnetic field increase in the ramp (panel a).  Slightly later, in the downstream direction, the magnetic field decreased by a factor of five (panel a) in the absence of large amplitude potential fluctuations (panel c). At this time the perpendicular electron temperature (panel d) remained unchanged.}
\end{figure}

The electron perpendicular and parallel temperatures also correlate with the magnetic field of panel \ref{fig:X}a. While this might suggest that the compression of the magnetic field plays a role in the electron heating, it is known from the adiabaticity plot of Figure~\ref{fig:adiabaticity} that this is not the case. Another example of the relative unimportance of the magnetic field for electron heating is given in Figure~\ref{fig:extra} in which the temperature increase (panel \ref{fig:extra}d) correlates with the electric field turbulence (panel \ref{fig:extra}c) and also with the magnetic field increase (panel \ref{fig:extra}a). 
However, slightly later, the magnetic field decreases by a factor of five and there is no change in the perpendicular electron temperature (panel \ref{fig:extra}d), so the magnetic field magnitude is not controlling the electron perpendicular temperature. The adiabaticity for this event is 2.24. It is also noted that large values of $\delta n/n$ and $\delta \phi /T_{e\perp}$ have been reported for a different event \cite{sundkvist2013a}.

\section{Discussion}
It is shown that neither anomalous resistivity nor magnetized electron flow are capable of explaining thermal electron heating at terrestrial bow shock crossings observed by the THEMIS satellite. Instead, the combination of strong electric field turbulence and the cross-shock electric field in the NIF are required to understand why;
\begin{itemize}
\item
The perpendicular electron temperature increase differs from the magnetic field increase by factors as great as three.
\item
The parallel temperature increase is much greater than can be provided by the parallel electric field.
\item
The perpendicular and parallel temperature increases across the shock are, on the average, equal and they increases simultaneously.
\end{itemize}

Of the 70 bow shock crossings in the study, 67 had some level of electric field fluctucations similar to those described above. 
At the rate of 128 samples/second, 75\% of the crossings had turbulence levels greater than 10~mV/m. 
There was a weak dependence of the $>10$~mV/m turbulence on the solar wind electron temperature and magnetic field, with large amplitude turbulence found in all six events having a solar wind temperature above 12~eV and 10 of 11 events having a solar wind magnetic field greater than 10~nT.

Because the events were chosen selectively and because the apogees of the THEMIS satellites were about 11 Earth radii compared to a typical bow shock location of $\sim15$ Earth radii, the analyzed events may not be typical of the bow shock physics for a randomly selected set of crossings over a wider range of altitudes. To test this possibility, bow shock crossings by the Cluster 2 satellite during January, 2003, were examined for evidence of electric field turbulence. (Cluster 2 transmitted 0-180 Hz bandwidth data at a lower data rate, so this aliased data was appropriate for observing whether turbulence was present, but not for quantitatively analyzing it.) The 140 clean Cluster bow shock crossings during this month occurred over radial distances of 12.9 to 18.9 Earth radii, and 90\% of them had fluctuation amplitudes $>10$~mV/m. Thus, the THEMIS events in the current study were not anomalous and the results of this study may be extrapolated to all terrestrial perpendicular bow shocks. 

We finally arrive at the following new scenario for thermal heating of electrons in super-critical shocks, consistent with the observations presented above. 
Instabilities due to gradients, currents and possibly unstable particle distributions in the shock ramp are driving strong nonlinear fluctuations, primarily in the electric field. When a magnetized electron drifts into regions of strong potential fluctucations on the scale of a gyro orbit, $\delta \phi/T_{e\perp} \sim 1$, the electrons demagnetize in this local region, 
$\delta \phi/T_{e\perp}$, $T_{e\perp}$, $T_{e||}$ and the DC electric field correlate and most of the electron heating occurs. 
This scenario is required because it has been shown that neither magnetized electron heating nor heating due to anomalous resistivity is consistent with the experimental data.
It should however be pointed out that such strong potential fields are likely to trap electrons, which could lead to secondary effects not explored here. 

Because astrophysical shocks should be stronger than terrestrial bow shocks, the mechanism of electron heating by the combination of strong electric field turbulence and a cross-shock electric field may be applicable to all of astrophysics.

\begin{acknowledgments}
This work was supported by NASA contract NAS5-02099-09/12 and NASA grants NNX13AE24G, NNX09AE41G-1/14. The authors acknowledge the considerable assistance of Dr. J. P. McFadden and Professor S. Bale in the collection and interpretation of the plasma data.
\end{acknowledgments}


\end{document}